\documentclass[aps,pra,twocolumn]{revtex4-1}

\usepackage{upgreek}
\usepackage{epsfig}
\usepackage{tikz}
\usepackage{graphicx}
\usepackage{epstopdf}
\usepackage{fancybox}
\usepackage{makeidx}
\usepackage{mathrsfs}
\usepackage{appendix}
\usepackage{amsmath}
\usepackage{amsfonts}
\usepackage{natbib}
\usepackage[figuresright]{rotating}
\usepackage{floatpag}
\usepackage{natbib}
\usepackage{marginnote}
\usepackage{enumerate}
\usepackage{hyperref}
\newcommand*{\myTagFormat}[2]{(\ref{#1}$#2$)}
\usepackage{color, colortbl}
\definecolor{LightCyan}{rgb}{.5,1,0}

\newcommand{\Zbb}{\mathbb{Z}_2}

\newcommand{\Rbb}{\mathbb{R}}

\newcommand{\Nbb}{\mathbb{N}}

\begin{document}

\title{MemComputing Integer Linear Programming}

\author{Fabio L. Traversa}
\email{email: ftraversa@memcpu.com}
\affiliation{MemComputing, Inc., San Diego, CA, 92130 CA}

\author{Massimiliano Di Ventra}
\email{email: diventra@physics.ucsd.edu}
\affiliation{Department of Physics, University of California, San Diego, La Jolla, CA 92093}




\date{\today}

\begin{abstract}
Integer linear programming (ILP) encompasses a very important class of optimization problems that are of great interest to both academia and industry. Several algorithms are available that attempt to explore the solution space of this class efficiently, while requiring a reasonable compute time. However, although these algorithms 
have reached various degrees of success over the years, they still face considerable challenges when confronted with particularly hard problem instances, such as those of the MIPLIB 2010 library. In this work we propose a radically different {\it non-algorithmic} approach to ILP based on a novel physics-inspired computing paradigm: Memcomputing. This paradigm is 
based on {\it digital} (hence scalable) machines represented by appropriate electrical circuits with memory. These machines can be either built in hardware or, as we do here, their equations of motion can be efficiently simulated on our traditional computers. 
We first describe a new circuit architecture of memcomputing machines specifically designed to solve for the linear inequalities representing a general ILP problem. We call these {\it self-organizing algebraic circuits}, since they self-organize dynamically to satisfy the correct (algebraic) linear inequalities. We then show simulations of these machines using MATLAB running on a single core of a Xeon processor for several ILP benchmark problems taken from the MIPLIB 2010 library, and compare our results against a renowned commercial solver. We show that our approach is very efficient when dealing with these hard problems. In particular, we find within minutes feasible solutions for one of these hard problems (f2000 from MIPLIB 2010) whose feasibility, to the best of our knowledge, has remained unknown for the past eight years. 
\end{abstract}    

                          
\maketitle
\section{Introduction}\label{Intro}

Integer programming represents an important tool to describe a variety of optimization problems that appear both in industry and academia~\cite{Schrijver1998}. The general format of integer programming consists of an objective function to be minimized over a set of variables and subjected to a set of constraints defined by linear inequalities among variables. In addition, the variables are constrained to be integer values.  If the objective function is linear, then we properly refer to Integer Linear Programming (ILP), which is the problem class we consider in this paper.

Due to its fundamental and practical importance, ILP is still extensively studied in both academia and industry. Several general-purpose open source~\cite{GLPK,Forrest2018,Gleixner2018,Ralphs2017,Berkelaar2004} and commercial solvers~\cite{Gurobi,CPLEX,FICO,MATLAB} have been developed together with specialized solvers specifically optimized for ILP, with additional structures usually developed for some specific projects~\cite{applegate2006concorde,Floudas2005,Boston99,sorensen2014hybridizing}.

Solutions to ILP can be approached via different algorithms, both heuristics~\cite{Danna2004,Boston99,Glover1977} and exhaustive~\cite{Schrijver1998,barnhart1998branch,benders1962partitioning,hooker2003logic}. Quite often, exhaustive methods are also called ``complete algorithms''. The complete algorithm for ILP that is most commonly employed is a 
combination of cutting planes and branch-and-bound, also known as the branch-and-cut algorithm~\cite{Schrijver1998}. All these solvers have demonstrated several degrees of success on a variety of ILP problems~\cite{abara1989applying,kroon2009new,stahlbock2008operations,melo2009facility,bard2003staff}, but they still struggle when faced with particularly hard problems such as those within the MIPLIB 2010 library~\cite{MIPLIB2010}.

In this paper, we present a novel and general purpose {\it non-algorithmic} approach to the solution of ILPs based on the {\it memcomputing paradigm} previously introduced by two of us (FLT and MD)~\cite{UMM}. This new approach, which stands for computing {\it with} and {\it in} memory~\cite{13_memcomputing}, cannot be classified as stochastic search, since it does not use a probabilistic scheme, nor a trial and error strategy. In addition, memcomputing does not employ educated guesses and known structures of the problem to define a set of instructions for a program to find solutions to the problem at hand. In other words, the memcomputing approach is not algorithmic~\cite{Di_Ventra2018}. 

On the contrary, a given problem is embedded into an electronic circuit (a possible realization of Memcomputing Machines (MM)~\cite{UMM,DMM2,Di_Ventra2018}) whose time evolution ultimately relaxes to a steady state (equilibrium) that expresses the solution of the original problem~\cite{DMM2,Di_Ventra2018}). If these circuits are properly designed to satisfy several mathematical properties (see~\cite{DMM2,Di_Ventra2018}), they efficiently converge to the solution of the given problem, and chaos or periodic orbits can be avoided~\cite{no-chaos,noperiod}. However, for the 
case of optimization problems, the method does not provide proof of optimality for a given solution, nor does it detect the infeasibility of a problem.

In order to approach ILP we have first designed novel MMs based on the concept of {\it self-organizing algebraic gates} (SOAGs) we introduce in Section~\ref{Mem_section}. SOAGs are the building blocks of the MM to solve ILP and are designed so that their dynamics self-organize towards the equilibrium that represents the solution satisfying the constraints of the ILP. 

We have used this approach to find solutions for a selection of hard benchmark problems from the MIPLIB 2010 library~\cite{MIPLIB2010}. In Section~\ref{ILP_section} we introduce the basic nomenclature for ILP. In Section~\ref{Mem_section} we discuss the memcomputing approach for this class of problems, and in Section~\ref{Results_section} we provide and discuss numerical results by simulating the corresponding MMs to solve several open problems from the MIPLIB 2010 library. We then compare our performances against a renowned commercial solver (Gurobi) and show the efficiency of our approach, which has been able to find within minutes feasible solutions for one of these hard problems 
(f2000 from MIPLIB 2010) whose feasibility, to the best of our knowledge, has remained unknown for the past eight years.

\section{Integer Programming Basics}\label{ILP_section}

Let us consider a restricted version of ILP for which we have only binary variables. In this case, the problem can be formalized as follows:
\refstepcounter{equation}\label{ILP}
\begin{align}
\min_{\{x_j\}} \sum_j&f_jx_j\tag*{\myTagFormat{ILP}{a}}\\
A_{eq}x=&b_{eq}\tag*{\myTagFormat{ILP}{b}}\\
A_{ineq}x\le& b_{ineq}\tag*{\myTagFormat{ILP}{c}}\\
x_j\in\Zbb \text{ for}&\text{ each } j\tag*{\myTagFormat{ILP}{d}}
\end{align}
where $x=\{x_1,...,x_n\}$ with $x_j\in \Zbb$ for any $j=1,...,n$, $f_j\in \Rbb$, $A_{eq}\in \Rbb^{m_{eq}\times n}$, $b_{eq}\in \Rbb^{m_{eq}}$, $A_{ineq}\in \Rbb^{m_{ineq}\times n}$ and $b_{ineq}\in \Rbb^{m_{ineq}}$ with $m_{eq}$ and $m_{ineq}\in \Nbb$. This is also known as {\it 0-1 linear programming} and it is one of the Karp's 21 NP-complete problems \cite{complexity_bible}.

A solution  $\bar x$ of the ILP \eqref{ILP} is an assignment to $x$ such that all constraints (\ref{ILP}$b$)--(\ref{ILP}$d$) are satisfied. The solution $\bar x$ is said to be sub-optimal if $\sum_jf_j\bar x_j\ge \min_{\{x\}}\sum_jf_jx_j$ where $x$ satisfies (\ref{ILP}$b$)--(\ref{ILP}$d$). We also define the \emph{objective}, $O$, of the problem \eqref{ILP} as $O=\min_{\{x\}}\sum_jf_jx_j$. 

A lower bound for the ILP objective can be calculated efficiently in most cases by solving the relaxation problem obtained by replacing the constraint (\ref{ILP}d) with 
\begin{equation}
x_j\in[0,1] \text{ for}\text{ each } j\tag*{\myTagFormat{ILP}{d'}}.
\end{equation}
It is easy to prove that the objective $O_{LP}$ of the linear programming problem (\ref{ILP}$a$)--(\ref{ILP}$c$), (\ref{ILP}$d'$) satisfies $O_{LP}\le O$.

Finally, given a lower bound $O_{lb}$ of \eqref{ILP} a gap from optimality can be defined. Some commercial solvers estimate this gap as
\begin{equation}
   \textrm{gap} = \dfrac{O_{best}-O_{lb}}{O_{best}}\label{gap_gurobi}
\end{equation}
where $O_{best}$ is the best objective found so far while $O_{lb}$ is the best lower bound to the objective found so far. Therefore in many cases $O_{LP}$ can be used as a lower bound for the gap if no better lower bound is available.

\section{MemComputing Strategy}\label{Mem_section}

The memcomputing approach to ILP problems is based on the concept of Self-Organizing Algebraic Gates (SOAGs). SOAG is a novel circuit design developed at MemComputing, Inc.~\cite{MemWeb} by one of the authors (FT). It is inspired by the previous work on Self-Organizing Logic Gates (SOLGs)~\cite{DMM2,patentSOLC}. 
Both SOLGs and SOAGs are building blocks for practical realizations of Universal Memcomputing Machines (UMM)~\cite{Di_Ventra2018,UMM,traversaNP}, in particular their {\it digital} (hence scalable) sub-set: digital memcomputing machines (DMMs)~\cite{DMM2}.

The main properties of SOLGs have been recently investigated and it has been proved that a proper design leads to Self-Organizing Logic Circuits (SOLCs) that demonstrate long-range order and topological robustness~\cite{topo,Bearden2018}. Moreover, SOLCs can be designed in such a way that persistent chaotic and oscillatory behavior can be avoided~\cite{no-chaos,noperiod}. SOLCs have also been proved to be very efficient in a variety of combinatorial optimization problems such as maximum satisfiability (MAXSAT)~\cite{Traversa2018,Sheldon2018}, quadratic unconstrained binary optimization (QUBO),  spin-glasses~\cite{spinglass}, and pre-training of deep-belief networks~\cite{AcceleratingDL}.     

\begin{figure}
	\centerline{\includegraphics[width=1.02\columnwidth]{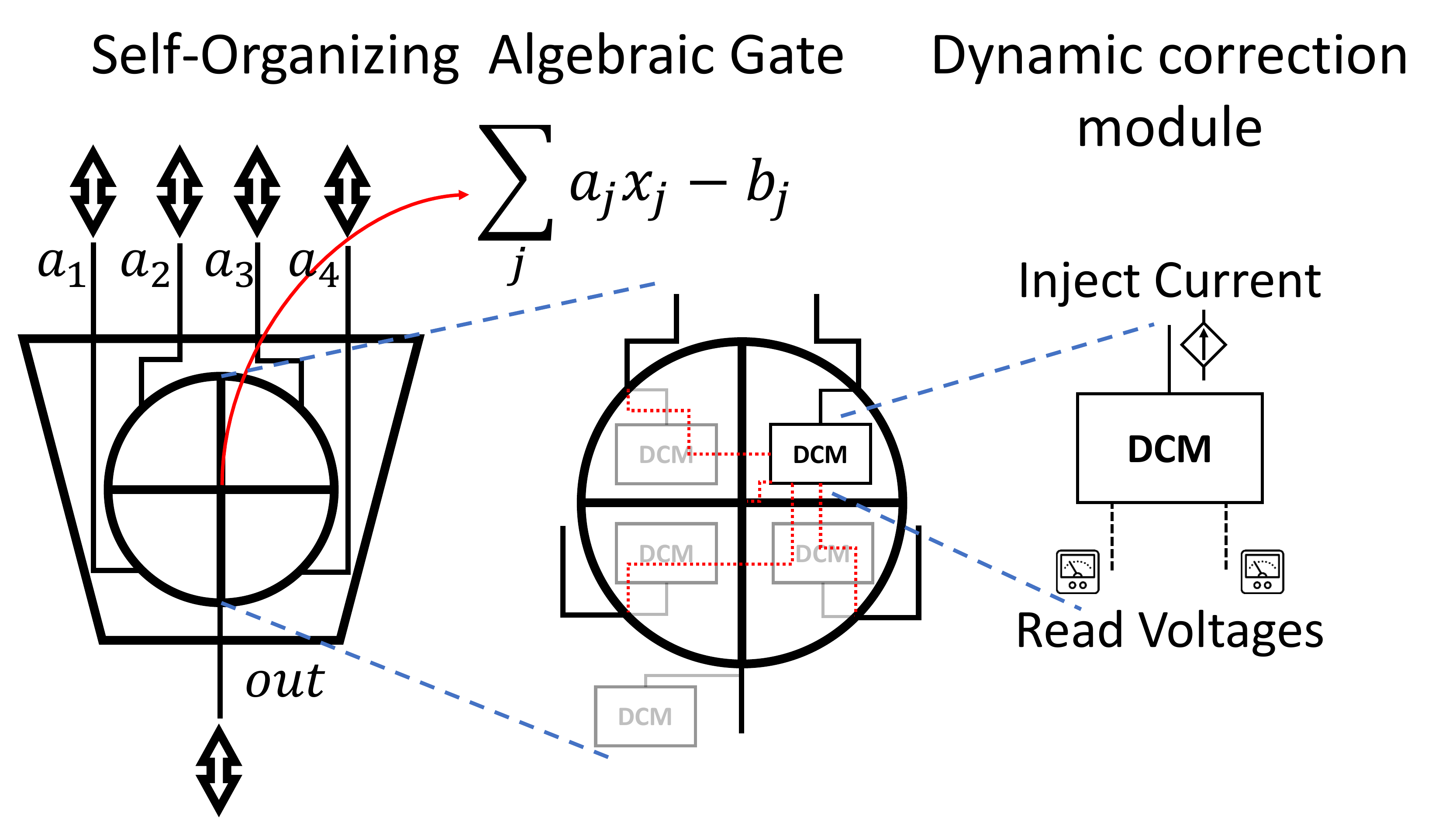}}
	\caption{Sketch of a Self-Organizing Algebraic Gate. All terminals allow a superposition of incoming and outgoing signals from the surrounding circuit. The central unit processes the signals in order to satisfy a liner algebraic relation consistent with the requirement of the ``out'' terminal. The self-organization is enforced by the Dynamic Correction Modules that read voltages from all terminals and inject a current to the appropriate terminal as long as the algebraic relation is not satisfied.}
	\label{figSOAG}
\end{figure}

SOLCs can be classified as {\it digital} realizations of UMMs because they accept inputs and return outputs that are digital in nature. Input and output are related to the circuit realization by associating logical 0s or 1s to voltages that are below or above a threshold, respectively. In this way, the required precision in writing inputs and reading outputs is {\it finite} and {\it independent} of the size of the problem at hand. However, the transition function of these machines 
(namely the function that maps input to output) is physical (analog) and takes full advantage of the {\it collective state} of the system to process information~\cite{collective,traversaNP,DMM2}. We reiterate though that, despite the physical nature of the transition function, DMMs can easily {\it scale}
 because they {\it do not} require precision that increases with the size of the problem. Rather, they can handle, ideally, unbounded problem sizes~\cite{Sheldon2018}.

SOAGs share the same principles and scalability advantages of SOLGs but their circuit is designed to self-organize toward an {\it algebraic relation} rather than a boolean relation as for SOLGs. In this work, the SOAGs have been designed to satisfy linear relations between boolean variables as a particular case of algebraic relations (see Fig.~\ref{figSOAG}). Further extensions of this design will include mixed integer and continuous variables, as well as nonlinear algebraic relations.    

\begin{figure}[t!]
	\centerline{\includegraphics[width=1.02\columnwidth]{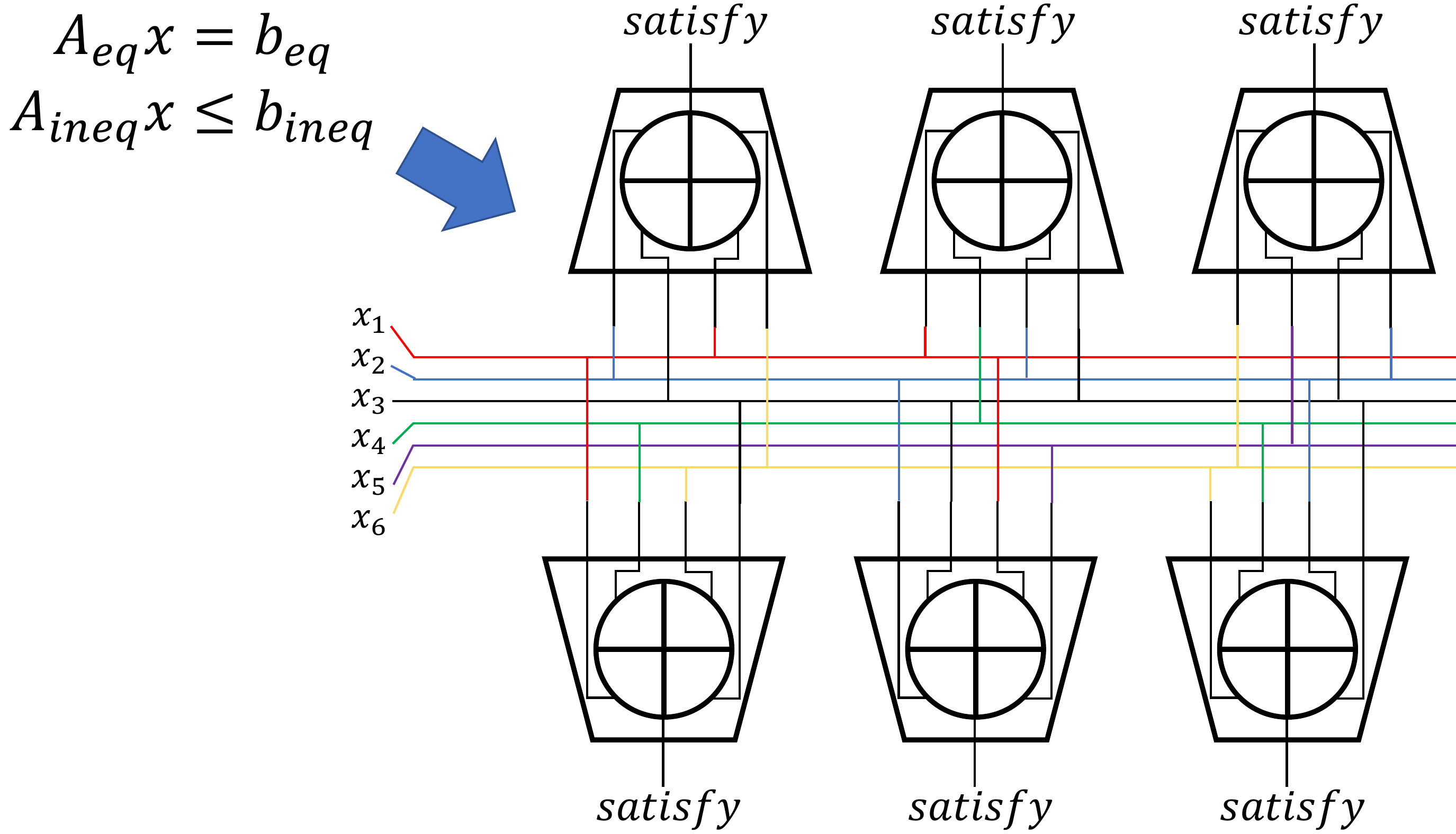}}
	\caption{Sketch of a Self-Organizing Algebraic Circuit (SOAC). SOAGs are connected together in an architecture that directly maps the ILP into the SOAC.\label{figSOAC}}
\end{figure}

By connecting together SOAGs, we then assemble a Self-Organizing Algebraic Circuit (SOAC), see Fig.~\ref{figSOAC}. The SOAC collectively self-organizes in order to satisfy the relations embedded in the gates. In this way, it is trivial to embed the problem~\eqref{ILP} directly into the SOAC. Each one of the equations in (\ref{ILP}b) and (\ref{ILP}c) is mapped directly into a SOAG, while (\ref{ILP}a) can be easily reformulated as an extra linear inequality 
\begin{equation}
\sum_jf_jx_j\le \tilde b\tag*{\myTagFormat{ILP}{a'}}.
\end{equation}
where $\tilde b$ is an extra parameter that can be dynamically changed in the circuit in order to find solutions of increasing quality, each time closer to the global optimum. Finally, (\ref{ILP}d) is naturally embedded in the circuit since inputs and outputs are digital. 

Like the SOLCs, the ultimate {\it physical} electrical circuit representing a SOAC contains active and passive elements, with and without memory (internal state variables)~\cite{DMM2,Di_Ventra2018}. 
The corresponding electrical circuit can be built with available complementary metaloxide semiconductor
(CMOS) technology. However, since its components are {\it non-quantum}, the ordinary differential equations describing it can be efficiently simulated 
on our modern computers. These equations are of the type 
\begin{equation}
\dot y = F(y), \;\;\;\;y(t=0)=y_0,
\end{equation}
with $y$ a vector describing all voltages/currents and internal state variables of 
the system, and $F$ the flow vector field describing its dynamics~\cite{DMM2,Di_Ventra2018}. These non-linear differential equations are then integrated numerically in time from a given (random) initial condition $y(t=0)=y_0$ up to a time out (TO) we set at the outset. Finally, the voltages of the state $y(t=\textrm{TO})$ represent 
the solution $\bar x$ of the ILP \eqref{ILP} at hand. Below, we present the results of these numerical simulations.

\section{Numerical Results}\label{Results_section}

We consider as benchmark instances, problems from the MIPLIB 2010 library \cite{MIPLIB2010}. In particular, we consider the class of {\it open problems}, and we focus on the 0-1 programming ones only. This represents a set of 24 benchmark problems. The open problems are classified as problems for which either the optimal solution has never been found, or proved to be optimal, or, in some cases, the feasibility of the problem is unknown. This benchmark is a unique collection of problems from several industries or competitions for solvers. It represents a standard benchmark that developers use to test their solvers. 
\begin{table}[th!]
	\begin{center}
		{
			\begin{center}
				\begin{tabular}{ |p{4cm}|p{2.5cm}|  }	
					\cline{1-2}
					\multicolumn{1}{|c|}{Parameter Name} & \multicolumn{1}{c|}{Parameter Value} \\
					\hline\hline
					TimeLimit	&	3380\\
					\hline
					Presolve	&	2\\
					\hline
					Method		&	3\\
					\hline
					MIPGapAbs	&	0\\
					\hline
					MIPGap		&	0\\
					\hline
				\end{tabular}
			\end{center}
		}
	\end{center}
	\caption{Gurobi 8.0 parameters used in the tests. The same values for Gurobi solver parameters were used across all models. The {\it TimeLimit} parameter was set to 3380 seconds to allow a 220 second buffer for cloud server booting and shutdown while staying within one hour of total machine time.  The {\it Presolve} parameter was set to 2 (maximum) in expectation that, more often than not, any additional presolve time would be offset by finding improved solutions earlier, given the known difficulty of the models.  The {\it Method} parameter was set to 3 (concurrent) to allow maximum use of the 16 threads at the root node of the model in addition to using all 16 threads by default beyond the root node.  The {\it MIPGapAbs} and {\it MIPGap} parameters were both set to 0 to ensure the Gurobi solver would not terminate at a suboptimal solution prior to the time limit. \label{Table_Gurobi_parameter}}
\end{table}

In order to compare the performance of the MemComputing ILP solver (we refer to it as ``MemCPU''), we have run these problems also using the Gurobi 8.0 solver~\cite{Gurobi}. Gurobi is a renowned commercial solver for mixed-integer programming (MIP) used worldwide. In most cases it is employed as a reference because of its high-quality performance. Gurobi is a complex agglomerate of algorithms and heuristics to improve the time to, and quality of the solution of mixed-integer programming problems~\cite{Gurobi_basic}. The main algorithm implemented in Gurobi is the branch-and-bound~\cite{computational_complexity_book}. However, the latter is boosted by pre-processing, including variable pre-solving, cutting planes, and heuristics. In addition, Gurobi employs sophisticated algorithms and heuristics in order to further accelerate the branch-and-bound procedure~\cite{Gurobi_basic}. The result is a collection of state-of-the-art algorithms, solution strategies and optimization toward the solution of MIP problems. 

We stress here once more the major difference between our MemCPU solver and Gurobi. The former solves {\it differential equations} of a 
{\it physical system} that represents the original ILP problem. Gurobi, instead, is a sophisticated but still traditional (combinatorial) {\it algorithmic} approach. In 
other words, the memcomputing approach {\it first} transforms the original optimization problem into a Physics problem, and {\it then} simulates the dynamics 
of such a physical system, while maintaining the digital structure of inputs and outputs~\cite{Di_Ventra2018}. 

\begin{table*}
	\begin{center}
\begin{tabular}{ |p{4cm}||p{2.5cm}|p{2.5cm}||p{2.5cm}|p{2.5cm}|  }	
	\cline{2-5}
	\multicolumn{1}{ c||}{} & \multicolumn{2}{c||}{MemCPU}  & \multicolumn{2}{c|}{Gurobi 8.0} \\
	\hline \hline
	\multicolumn{1}{|c||}{File Name}  & \multicolumn{1}{|c|}{300s}  & \multicolumn{1}{c||}{3380s}   & \multicolumn{1}{c|}{300s}  & \multicolumn{1}{c|}{3380s} \\
	\hline\hline
bab1	&	-197710.06 &	-202800.20&	-218764.89&	-218764.89\\
	\hline
bab3	&	TO&	TO&	-654569.46&	-656193.04\\
	\hline
circ10-3&		362.00&	312.00&	TO&	386.00\\
	\hline
datt256&		TO&	TO&	TO&	TO\\
	\hline	
ds-big	&	29780.20&	6902.00&	762.93&	762.93\\
	\hline
ex1010-pi&		237.00&	237.00&	240.00&	238.00\\
	\hline
f2000	&	1950.00&	1846.00&	TO&	TO\\
	\hline
ivu06-big&	TO	&	349.02&	9416.00&	159.96\\
	\hline
methanosarcina&		2734.00	&2731.00&	2756.00&	2737.00\\
	\hline
neos-952987&		TO&	TO&	TO&	TO\\
	\hline
ns1853823&		144000.00&	84000.00&	284000.00&	124000.00\\
	\hline
ns894236&		17.00&	17.00&	17.00&	17.00\\
	\hline
ns894786&		13.00&	13.00&	14.00&	13.00\\
	\hline
ns903616&		20.00&	19.00&	20.00&	19.00\\
	\hline
pb-simp-nonunif&		87.00&	71.00&	75.00&	42.00\\
	\hline
ramos3&		186.00	&186.00&	252.00&	244.00\\
	\hline
rmine14&		-4208.27&	-4223.09&	-194.06	&-4283.04\\
	\hline
rmine21	&	-9340.90&	-10392.05&	TO&	-214.18\\
	\hline
rmine25	&	-13295.47&	-15037.29&	TO&	-185.33\\
	\hline
sts405	&	340.00&	340.00&	342.00&	342.00\\
	\hline
sts729	&	617.00&	617.00&	650.00&	648.00\\
	\hline
t1717	&	206003.00&	192840.00&	201342.00&	201342.00\\
	\hline
t1722	&	126537.00&	119071.00&	129822.00&	119764.00\\
	\hline
zib01&		TO&	TO&	TO&	TO\\
	\hline
\end{tabular}
	\end{center}
\caption{Objectives found after running MemCPU and Gurobi 8.0 for 300 seconds and 3380 seconds. Objectives are in arbitrary units and TO $=$ Time Out. The problems are from the class ``open'' of MIPLIB 2010 library~\cite{MIPLIB2010} restricted to the 0-1 programming only.  \label{Table_runs}}
\end{table*}

\begin{figure}[t!]
	\centerline{\includegraphics[width=1.02\columnwidth]{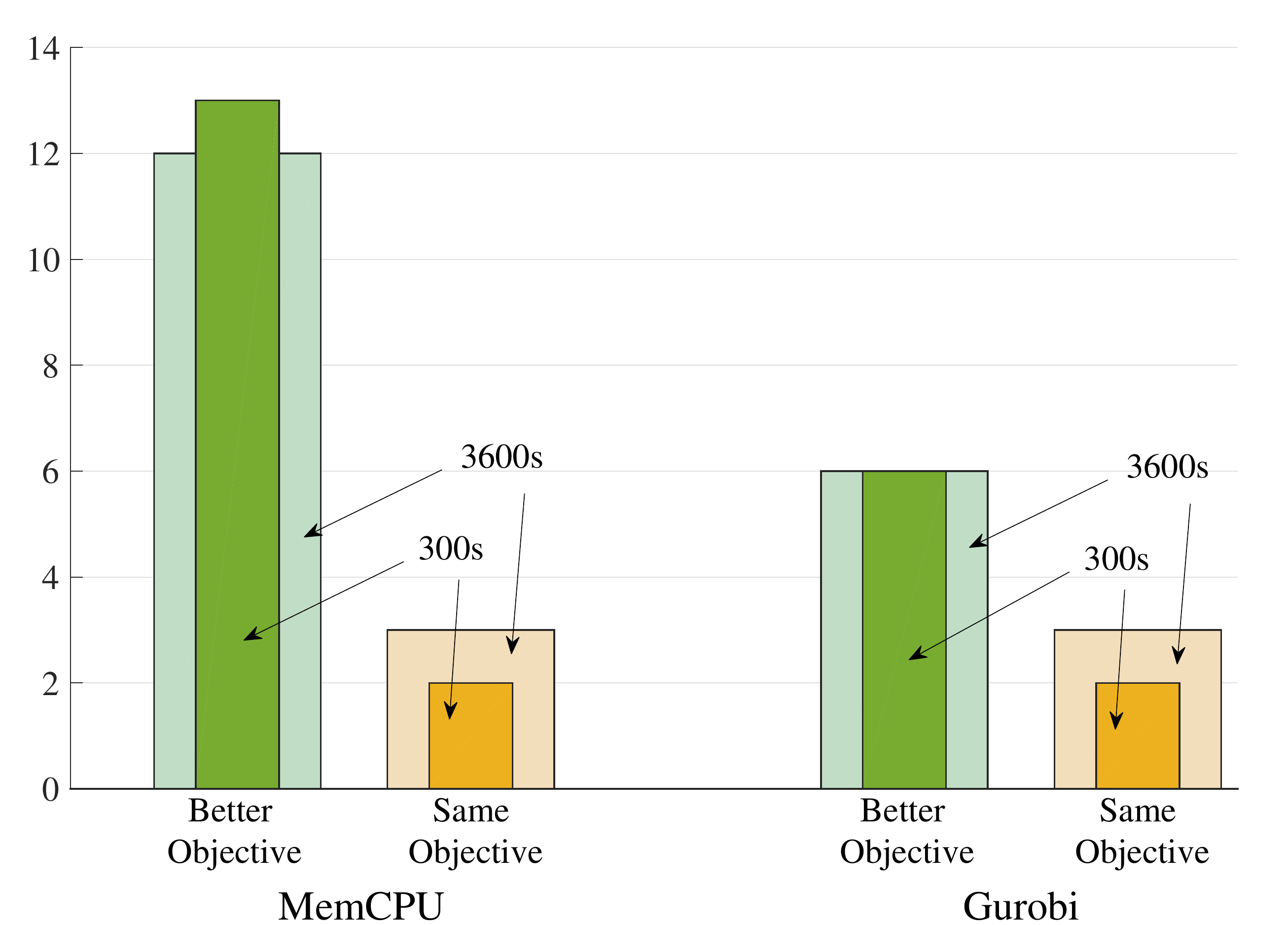}}
	\caption{Histogram of better objectives found by MemCPU or Gurobi 8.0 from Table~\ref{Table_runs}. The problems for which both solvers did not find any feasible solution have been excluded. \label{figHist}}
\end{figure}

We have compared MemCPU versus Gurobi 8.0 in two different regimes: 300 seconds and 3380 seconds total running time. The Gurobi Solver was run on the Gurobi Cloud using an Amazon Web Services c4.4xlarge instance having 16 CPU cores and 30~GB RAM and the settings are described in Table~\ref{Table_Gurobi_parameter}

The MemComputing, Inc. solver, MemCPU, has been implemented in interpreted MATLAB and run on an Intel Xeon 6148 with 192~GB RAM using only up to 10 cores at a time. However, the multi-core processing has only been used to run up to 10 identical versions of the solver ({\it replicas}) in parallel using {\tt parpool}. The {\tt parpool} function guarantees that each replica does not use multi-threading. Therefore, each replica used exactly one core. 

Each problem has been processed by MemCPU generating a SOAC representing the ILP problem without any pre-processing or variable pre-solving. Since MemCPU simulates an electronic circuit, there are physical parameters to be set that accelerate the self-organizations of the circuit depending on the problem at hand. Therefore, these parameters needed to be tuned~\cite{tuning}. The replicas have been used to run MemCPU with different initial conditions and different parameter sets for tuning purposes. The best outcome from the replicas has been selected as output.

In Table~\ref{Table_runs} the objectives after 300s and 3380s runs are reported for both MemCPU and Gurobi 8.0 using the aforementioned Gurobi parameter set. In Fig.~\ref{figHist} there is a direct comparison between MemCPU vs Gurobi by counting the number of problems for which each solver found solutions with better objective function values than the other. From the overall comparison we can see that the direct approach of MemCPU to solving ILP resulted in finding solutions with better objective function values on more than twice as many problems.

In the following subsections we discuss in detail some relevant results and characteristics of the MemCPU solver. 

\subsubsection{The {\tt f2000} Problem}\label{f2000}

One very interesting result that deserves to be discussed separately is related to the outcome of MemCPU for the {\tt f2000} problem of the MIPLIB 2010 library. The {\tt f2000} problem belongs to the class of hard random problems~\cite{computational_complexity_book} and was selected from the pseudo-boolean competition 2010~\cite{pseudoboolen2010}, that was a special event of the satisfiability (SAT) 2010 conference. 
Since then, {\tt f2000} has been part of the next editions of the competition and also part of the open problems of the MIPLIB 2010 library~\cite{MIPLIB2010}. 

Despite many groups from both SAT and MIP communities having tried, using both complete and incomplete solvers, to find feasible solutions for this problem during the past eight years, to the best of our knowledge, no one has been able to find a feasible solution yet. Currently the feasibility of {\tt f2000} is classified as ``unknown'' by MIPLIB~\cite{MIPLIB2010}. 

Running this problem using the MemCPU solver, we already found, within 60s, the first feasible solution to the problem, and for longer run times more solutions with objectives of increasing quality (see also Sec.~\ref{data}). This result is then representative of both the uniqueness and power of the memcomputing approach.

\subsubsection{Deep Diving Objectives}\label{Deep}

Looking closer at Table~\ref{Table_runs}, we can see that for many problems, irrespective of their size and structure, MemCPU found very quickly much better objectives. This shows that, for these problems, solutions with objectives much closer to the global minimum are strong attractors for the SOAC. For some of them, the convergence was so quick that Gurobi did not find in one hour what MemCPU found in five minutes, possibly demonstrating the capacity for orders of magnitude speed-up of the memcomputing approach versus the traditional algorithmic one. 

For some of the problems (e.g., {\tt ramos3}, {\tt ns1853823}, {\tt ex1010-pi}), it is also possible that MemCPU may have found the global optimum since no refinement in the objective was found after minutes of run time. However, since MemCPU does not provide proof of optimality, we could not prove mathematically the validity of this statement.

Finally, it is worth mentioning that for the problems for which Gurobi performed better than MemCPU, from its log file (and also comparing the results for 300s and 3380s runs) it is clear that Gurobi's pre-processing was particularly effective at simplifying the problem which may have contributed to it finding a very good initial feasible solution. Instead, as we have already discussed, MemCPU does not include any pre-processing and it starts from random initial conditions. In addition, for these problems, both the MemCPU and Gurobi parameter tuning may not necessarily be optimal. At the moment, MemComputing, Inc. is working on an automatic routine to predict and fine-tune the MemCPU parameters ~\cite{tuning}. Furthermore, since the design of SOAGs is not unique, further improvements that may accelerate the simulations and provide better solutions may be possible. Work along these lines is underway.     

\subsubsection{Scaling with Problem Size}\label{scaling}

\begin{figure}[t!]
	\centerline{\includegraphics[width=1.02\columnwidth]{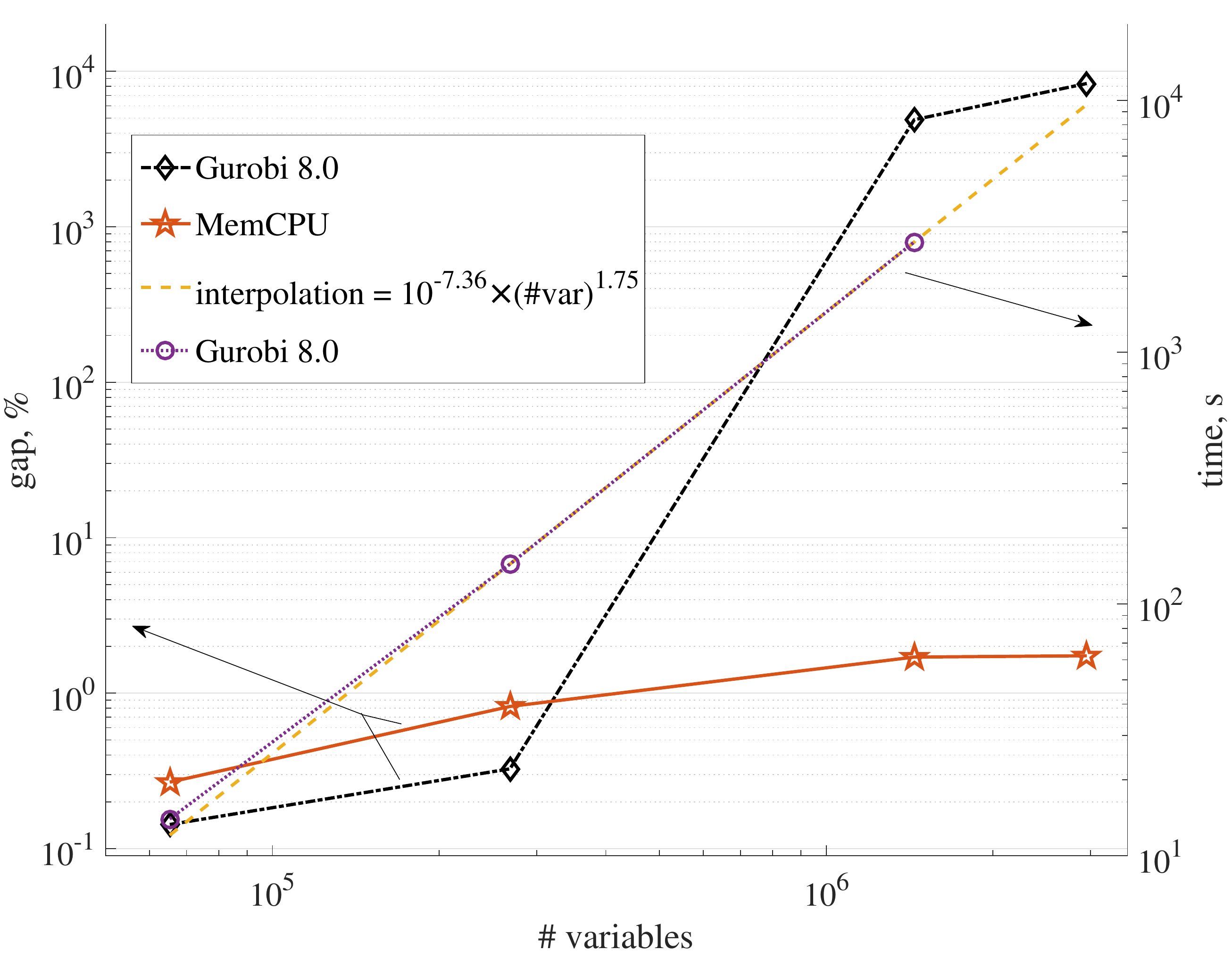}}
	\caption{Left y-axis: The gap defined in Eq.~(\ref{gap_gurobi}) for the {\tt rmine} benchmark for both MemCpu and Gurobi 8.0. Right y-axis: Gurobi 8.0 pre-processing time.  \label{figscaling}}
\end{figure}

The {\tt rmine} benchmark is a series of problems that model the open pit mining problem~\cite{MIPLIB2010}. This problem is industrially relevant and has been heavily studied by the MIPLIB organizers themselves in the past years~\cite{Shinano2016}.  Moreover, the MIPLIB 2010 library contains 5 instances with increasing numbers of variables depending on the refinements used to represent the problem. Therefore, this particular problem provides a benchmark for studying the scaling properties of both solvers. 

We have then considered runs of 3380 seconds to assess the scaling properties for this benchmark. However, we include in the analysis only 4 out of the 5 instances because the smallest one ({\tt rmine6}) belongs to the ``easy'' category. Therefore, it is exactly solved in less than 3380s. 

In Fig.~\ref{figscaling} we report the gap defined in Eq.~(\ref{gap_gurobi}) for both Gurobi 8.0 and MemCPU after 3380s run. Gurobi converges very fast for the small instances to objectives smaller than 1\%. However, by increasing the number of variables (namely, by increasing problem refinement), 
Gurobi was unable to complete the root simplex/barrier within the time limit and, consequently, showed much worse performance, providing solutions with objectives on the order of tens of thousands of \%. On the other hand, MemCPU, even though, for smaller instances, had a slightly slower convergence to find solutions, it maintained the scaling also at high numbers of variables, providing a solution at about 1\% gap for very large instances as well.   

In order to understand better the reason for this difference between MemCPU and Gurobi, from the log file of the latter we found that the pre-processing time for those instances grew almost quadratically as shown in Fig.~\ref{figscaling}. We acknowledge that this could be largely a consequence of having set the {\it Presolve} parameter to its maximum value. However, the pre-processing is only the starting point of the solution process used by Gurobi that for these problems can simply grow unbounded. This is easily seen, for example in the {\tt rmine21} problem, for which Gurobi was able to finish the pre-processing in a reasonable amount of time, but then spent 39 minutes on the root simplex/barrier, and then no refinement of the gap was made in the remaining time by the branch-and-bound process.

\subsubsection{Data Availability}\label{data}

In order for the interested reader to check all the results discussed in the previous sections, and confirm the validity of what we have reported in this paper, we have made available all solution files on MemComputing, Inc. webpage: {\tt \url{http://memcpu.com/downloads}}. The files are in the .sol format, and each file name corresponds to the name of the problem that can be downloaded from the MIPLIB 2010 library~\cite{MIPLIB2010}.

By using these data files the reader can use any solver for ILP (for instance Gurobi), and verify that all the objectives found by MemCPU and reported in Table~\ref{Table_runs} are indeed feasible solutions.

\section{Conclusions}\label{Conclusions}
In summary, we have shown how to employ {\it digital} (hence scalable) memcomputing machines to tackle the important problem class of integer linear programming problems. We have proposed a new set of self-organizing gates that self-organize to satisfy algebraic relations. When assembled together, these gates form a self-organizing circuit specifically designed to solve a given ILP problem. 

We have then simulated the corresponding equations of motion of these circuits to find solutions to a variety of benchmark ILP problems as reported in the MIPLIB 2010 library. We have compared our results with a well-known commercial solver (Gurobi). 
Our solver is extremely efficient in finding very good objectives for these problems. 

In particular, we have found within minutes feasible solutions for the f2000 ILP problem (of MIPLIB 2010) whose feasibility, to the best 
of our knowledge, has remained unknown for the past eight years. We have also shown that our approach maintains a high quality of solutions with increasing size of the problem. 

Since our solver has been implemented using interpreted MATLAB, there is plenty of room to speed up the reported calculations. In addition, since memcomputing machines employ non-quantum systems, they can be easily implemented in hardware using standard electronic components, thus offering a realistic path to real-time computing for these, and other, important combinatorial optimization problems.


\bibliographystyle{naturemag}
\bibliography{SUSYref}

\end{document}